\documentclass[aps,preprint]{revtex4}

\begin{document}

\begin{titlepage}

 \noindent {\bf Weighted-residual methods  for  the solution
   of  two-particle Lippmann-Schwinger equation  without partial-wave decomposition}

  \begin{center}
  Zeki C. Kuruo\u{g}lu\\
  {\it Department of Chemistry, Bilkent University,
 06800 Bilkent, Ankara,Turkey}
  \end{center}

 Recently there has been a growing interest
 in computational  methods
 for quantum scattering equations
 that avoid the traditional decomposition of wave functions and scattering
 amplitudes into partial waves. The aim of the present work is to
 show that the weighted-residual approach  in combination with local basis functions
 give rise to convenient computational schemes
  for the  solution  of the multi-variable
   integral equations  without the  partial wave expansion.
   The weighted-residual approach provides a unifying framework
    for various variational and degenerate-kernel methods
    for integral equations of scattering theory.
    Using a direct-product basis
    of localized quadratic  interpolation  polynomials,
     Galerkin, collocation and Schwinger variational realizations of the
      weighted-residual approach   have been implemented
     for  a model potential. It is demonstrated that,
     for a given expansion basis, Schwinger
     variational method exhibits better convergence with
     basis size than Galerkin and collocation methods. A novel
     hybrid-collocation method is implemented with promising
      results as well. \\


\end{titlepage}

\begin {center} {\bf I. INTRODUCTION } \end{center}

The standard  approach to two-body scattering problem  has been
through angular momentum decomposition. This is advantageous for central potentials
as the equations for partial waves
decouple. However, for non-central potentials, partial waves are coupled and advantages
of partial wave expansion dispappear to a large extent.
Recently there has been a growing interest in computational
methods for quantum scattering equations
 that avoid the traditional decomposition of wave functions and scattering
 amplitudes into partial waves [1-11]. The reasons for this interest are many fold:
 At intermediate and
 high collision energies partial wave expansion
 are known to   converge very slowly. For example,
  in ion-atom  collisions number of partial waves necessary for convergence
  in the high energy regime  may run
  up to  several thousand [6, 10].  It appears that for most potentials, the
  scattering amplitudes are smooth, but
   partial wave amplitudes may show oscillatory behavior.
   Similarly, the off-shell two-body $T$-matrix
   has usually simple structure whereas partial wave components
   might strongly oscillate.
   Under such circumstances,  the partial wave expansion
   may be inadequate or even unreliable.
    In the context of three and four particle problems  to which the two-particle
   $T$-matrices are  the input, two-particle $T$-matrices requiring
   an excessively  large  number of partial waves would render even
   the angular momentum algebra
   too complicated and computationally difficult to perform [2].

      These observations suggest that, to treat two-particle scattering at high
      energies
      and  within the context of few-particle dynamics,
      one should work directly with vector momenta without resorting to
      expansions over angular momentum states. Significant progress has been reported
       on the formal and computational aspects of solving
       the three-particle Faddeev equations directly
       in terms of vector momenta [2, 13, 14].

  Thanks to the present day computing power,
  direct  numerical solution of three-dimensional
  Lippmann-Schwinger (LS) equation
   for two-particle $T$-matrix
   without recourse to partial wave expansion is within reach.
   The most straight forward approach is the so-called
   Nystrom method[15] in which  the integral equation is converted to
    a system of linear equations by approximating
    the multi-dimensional integral by a quadrature.
    Denoting with $(q,\theta, \phi)$
   the spherical components of
   the momentum vector ${\bf q}$,
   the dimension of the matrix problem would be $N_qN_{\theta}N_{\phi}$,
    if a direct-product quadrature scheme is used.  Here $N_q\, ,\, N_{\theta}\, $ and $N_{\phi}$
     are the number of quadrature points for the variables $q,\theta$ and $\phi$,
     respectively. The matrix size of this three-dimensional Nystrom method,
     however, might quickly get
     prohibitive and may require special computational  environment.
     Fortunately, however, for central potentials,
     dependence on the azimuthal angle $\phi$
     can be eliminated from the three-dimensional LS equation [1].
     This leads to an integral equation in two variables,
     solution of which can be carried out
     via the Nystrom method routinely
     in commonly available computational platforms.

     In three- and four-particle contexts, the two-particle
     $T$-matrix $<{\bf q}|T(E)|{\bf q}'>$  is needed
     at very many different two-particle energies $E$ and
     for a great many different off-shell momenta ${\bf q}$
     and ${\bf q}'$.
      Nystrom solutions may not be the most economical ways
      of generating the needed $T$-matrix elements.
      The aim of the present work is to
      show that the weighted-residual approach [17-20]
      in combination with a direct-product basis of local
      functions  provide
       efficient computational schemes for
      the two-particle transition matrix elements  $<{\bf q}|T|{\bf q}'>$
       in the form of separable    expansions.
       Generation of  off-shell  $T$-matrix elements via these separable expansions
      involves a matrix inversion   problem of much smaller order
      than the one encountered in the Nystrom   method.

          The weighted-residual idea (also known as the Petrov-Galerkin approach) [17-20]
          provides  a unifying  framework for basis-set expansion methods
          for the solution of differential and integral equations.
      It involves  two function spaces: a finite-dimensional
     approximation subspace (of trial solutions) and a subspace  of
     test (or weight) functions. The coefficients of the expansion
     of the unknown function
     over the approximation subspace are determined
     by requiring  the residual (i.e., the difference between
      the exact and approximate solutions)
     to be orthogonal to the test space.
     For the two-particle LS equation this
     leads to a separable expansion of the T-operator.
     We discuss the connections of
     this separable expansion with oblique projections [21,22,23] and
     inner projections [24,25].
     Depending on the choices made for
     the approximation and test spaces,
      the weighted-residual approach give rise to a wide range of methods.
      Galerkin method, collocation method  and
      method of moments are the well-known examples.

          Schwinger variational (SV) method  represents  another
          instance of the weighted-residual approach.
           Of course, the Schwinger variational principle
           is well known  and has been widely used to solve
         partial-wave (single-variable)  LS equations.
        Reference [16] gives a comprehensive review of ( and
        an extensive list of references for)
        the formal and computational aspects
        of Schwinger variational methods.
         The present paper  demonstrates
         that Schwinger variational method (SVM) is
         a versatile tool
         for the  solution  of the vector-variable LS equations
         without the partial wave expansion as well.
           In particular, we show that,  for the potential
           considered in this work, SVM exhibits better
            convergence with basis size
            than the  Galerkin and collocation methods.

           We note that Galerkin method in conjunction with
           a direct-product basis of wavelets
           has been used in Ref.[5]
            to solve the two-dimensional LS equation
            with  a model two-nucleon  potential.
            It is likely that the two-dimensional wavelets
            will also prove efficient
            if used as the expansion basis in SVM.
            Another application of the Galerkin method
             without partial wave expansion has been made in Ref. [3] to
             solve the two-variable Schrodinger equation
             subject to scattering boundary conditions
             using a direct-product basis of local
             fifth degree polynomials.

             Plan of this article is as follows: In Sec. II, we discuss the reduction of  the
              three-dimensional  Lippmann-Schwinger equation
            into a two-dimensional integral equation. Some features
             of the reduced $T$-matrix is noted. Sec. III gives an exposition of the general
             weighted-residual approach in  the context of the Lippmann-Schwinger equation.
              The connection between the weighted residual method and  a projection approximation
              of the potential is established using the concept of  {\it oblique} projector.
              Sec. IV  discusses the various choices for the expansion and test spaces that lead to
               Galerkin, collocation  and Schwinger variational
               methods. A novel version of the collocation method
               (termed as {it hybrid}-collocation) that combines the
               advantages of  collocation and SV methods is
               formulated.
               Details of the computational
               construction of the expansion
               and test bases are described in Sec. V.
               The subtraction procedure to handle
               the singular integrals that come up
               in Petrov-Galerkin and Nystrom methods are
               discussed in this section as well. In Sec. VI   the
               results of calculations for a model potential are
               discussed and compared for different bases and methods.
               In Sec.VII we summarize our conclusions.

    \begin {center} {\bf II. LIPPMANN-SCHWINGER EQUATION } \end{center}

Basic equation for the description of two-particle scattering is
the Lippman-Schwinger equation for the two-particle transition
operator $T(z)$:
\begin{equation}
T(z)\ =\ V\ + V\, G_0(z)\, T(z)\ ,
\end{equation}
 where $V$ is the interaction potential between two particles,
 $G_0\,=(z-H_0)^{-1}$, with  $z$  being the (complex) energy of the two-particle system.
  For on-shell scattering,  $z=E+i0$ with $E=q_0^2/2\mu$.
  Working in  the center-of-mass frame, the eigenstates of the free Hamiltonian $H_0$
  are the relative momentum states $|\bf q>$.
  The matrix elements $T({\bf q},{\bf q}_0\, ; z)=
  <{\bf q}|T(z)|{\bf q}_0>$
  satisfy  the three-dimensional integral equation
 \begin{equation}
 T ({\bf q},\, {\bf q}_0 \, ;z)\ =\ V\, ({\bf q}, \, {\bf q}_0 )\, + \,  \int \, {\mbox{d} }{\bf q}' \,
 \frac{ V \, ({\bf q},{\bf q }' ) \ T ({\bf q}', {\bf q}_0 \, ; z ) } { z\,  -\, q^{'2}/2\mu }
\end{equation}
where $\mu$ is the reduced mass. Atomic units will be used throughout this article.
The $z$-dependence of $T$-matrix
elements  $T ({\bf q},\, {\bf q}_0 \, ;z)$ will   be suppressed in
the rest of this article. The momentum-space matrix elements $V\,
({\bf q}, \, {\bf q}' )\,$
 of the potential $V$ are given as
 \begin {equation}
 V ({\bf q}, \, {\bf q}' )\,= \ <{\bf q}|V|{\bf q}'> \ = \,  \int \, d{\bf r}\, <{\bf q}|{\bf r}>\, V({\bf r})
 \, <{\bf r}|{\bf q}'> \, ,
 \end{equation}
 with $<{\bf r}|{\bf q}> \, =\,  e^{i {\bf r}\cdot {\bf q} }/(2\pi)^{3/2}$.

 As first noted in Ref. [1],
  the azimuthal-angle dependence in Eq. (2)
  can be integrated out
  to obtain a two-dimensional integral equation.
  This is possible because
  $V ({\bf q}, \, {\bf q}' )$ and
 $T({\bf q}, \, {\bf q}' )$  in the case of central potentials depend
 only on $q,q'$ and $x_{qq'}= {\hat {\bf q} }\cdot {\hat {\bf q}}' = \, cos \,
 \theta_{qq'}$.
 Here $\theta_{qq'}$
 is  the angle between ${\bf q}$ and ${\bf q}'$ vectors.
  Denoting
 the polar and azimuthal angles of the momentum vector ${\bf q}\, $
 by $\theta  \, $
 and $\phi\, $,   respectively, we have
 $x_{qq'}=xx'+ ss'\, cos\, ( \phi-\phi')$,
  where $x=cos \, \theta $ and $s={\sqrt {1-x^2}}$.
 Whenever we want to make the functional
 dependence explicit, the notation
  $T(q,q', x_{qq'})$ will be used in place of
  $T({\bf q}, {\bf q}'\,)\, =\, T(q,\theta,\phi,\, q',\theta', \phi')\, $.

   We now introduce the reduced quantities
  \begin{eqnarray}
  V(q,x;q',x')\, =\, \int_0^{2\pi} \, d\phi\,\,  V({\bf q},{\bf q}')\,
  =\, \int_0^{2\pi} \, d\phi\,\,  V(q,q',x_{qq'})\, ,\\
  T(q,x;q',x')\, =\, \int_0^{2\pi} \, d\phi\, \,  T({\bf q},{\bf q}')\,
  \, =\, \int_0^{2\pi} \, d\phi\, \,  T(q,q',x_{qq'})\,.
  \end {eqnarray}

  \noindent The crucial observation [1] is that the above integrals
  are independent of the value of
  the azimuthal angle $\phi'$. In fact, if we define an averaged momentum
  state $\, |qx>\, $ as
  \begin {equation}
  |qx>\,\,  =\,  (2\pi)^{-1/2}\, \int_{0}^{2\pi}\, d\phi \,\, |{\bf q}>  \,
  \end{equation}
  one can easily verify that $\, V(q,x;q',x')\,\,  =\,\,  <qx|V|q'x'>\,\,  $ and
   $\, T(q,x;q',x')\,\,  =\\
    <qx|T|q'x'>\ $. This observation allows us to integrate Eq. (2) over $\phi$ and
  obtain  the two-dimensional Lippmann-Schwinger (LS2D) equation:
  \begin{equation}
  T(q,x;q_0,x_0)\, =\, V(q,x;q_0,x_0)\, + \ \int_{0}^{\infty}\,  dq'\, q'^2\, \int_{-1}^{1}\,  dx'
  \ \frac {V(q,x;q',x')\ T(q',x';q_0,x_0) } {z-q'^2/(2\mu)}\ .
  \end{equation}

  For an initial momentum vector ${\bf q}_0$
  along the z axis and a general final momentum vector $\bf q$,   we have
  $x_0=1$ and $x_{qq_0} =x$. Using this in Eq.(5), we find that
  \begin {displaymath}
  <{\bf q}|T|q_0\hat {\bf z}>\,  =\, T (q,q_0,x)\,
   =\,  (2\pi)^{-1}\,  T(q,x\, ;q_0, 1)\, . \nonumber
  \end{displaymath}
  Note also  that, when  $x_0=-1$, we have $x_{qq_0}=-x$ and
   $\ <{\bf q}|T|q_0\,  {\widehat {\bf -z} }>\,=\, T(q,q',-x)\,  =\, (2\pi)^{-1}\, T(q,x\, ;q_0,-1)\, $.
   It also follows from Eq.(5) that
   \begin{equation}
  T(q,x;q_0,x_0)\, =
  \, (2\pi)^{-1}\, \int_{0}^{2\pi}\, d\phi \ T(q,x_{qq_0};q_0,1),
  \end{equation}
  \noindent a relationship that might be useful in testing  the adequacy
  of numerical procedures employed to obtain  $T(q,x;q_0,x_0)\, $.

  The LS2D equation  can be solved by the Nystrom method in which
  the integrals over $q'$ and $x'$ are approximated
  by suitable quadrature rules and then   $x$ and $q$ variables are
  collocated at the quadrature points. We use the Nystrom method
  to obtain benchmark results against which the
   performance of the weighted-residual methods for different choices of
   expansion and test bases are tested. Computational implementation  of
  the Nystrom method is  outlined in Sec. V.

\begin {center} {\bf III. WEIGHTED-RESIDUAL APPROACH } \end{center}

   The weighted-residual methods
   (also known as the Petrov-Galerkin approach)
    involve a finite dimensional
   approximation subspace $\cal S_A$ (of trial solutions)
   and a subspace $ {\cal S_T}$ of
  test (or weight) functions.
  These two spaces are in general different,
  but are usually taken to have the same dimension.
  (The possibility exist, however,
  for using $\cal S_A$ and $\cal S_T$
  that have  have different dimensions,
  but the solution of the ensuing weighted-residual equations
   would then require
  the use of generalized-inverses.
  This is a possibility that we do not pursue in this
  article). If $\cal S_A$ and $ {\cal S_T}$ are taken to coincide,
  the resulting methods are referred to as Galerkin
  methods. Use of  different test subpaces $\cal S_T$
  with a given approximation space $\cal S_A$
  gives rise to a wide range of  Petrov-Galerkin methods
   (such as collocation , method of subdomains,
   least squares, and method of moments)[17-20]. The only compatibility
   requirement on the $({\cal S_A}, {\cal S_T})$ pairs
   is that no member of $\cal S_A$ be orthogonal to $\cal S_T$.

  The basis set for the approximation subspace $\cal S_A$ will be denoted as
  $\{\varphi_k(q,x)\, , \,  k=1,2,...,K\, \}$.
  The basis functions  are linearly independent,
  but not necessarily orthonormal.
   The projection operator onto the approximation subspace
   $\cal S_A$  is given  as
  \begin{equation}
  {\cal P_A}\, =
  \, \Sigma_{k=1}^{K}\,  \Sigma_{k'=1}^{K}\,
  |\varphi_k>\,  ({\bf \Delta}_{\cal A}^{-1})_{k,k'}\, <\varphi_{k'}| \, ,
  \end{equation}
  \noindent where ${{\bf \Delta}_{\cal A}}$ is the overlap matrix,
  viz., $({\bf \Delta}_{\cal A})_{k, k'} \, =\,   <\varphi_k|\varphi_{k'}>\, $.

  The test subspace $ {\cal S_T}$ is similarly spanned by
  a set of  linearly independent  functions
  $ \{ \chi_k (q,x)\, , \, k=1,2,...,K\,   \}$.
  The projection operator onto the test space is given  as
  \begin{equation}
  {\cal  P_T}\, =\, \Sigma_{k=1}^{K}\,  \Sigma_{k'=1}^{K}\,
   |\chi_k> ({ \bf \Delta} _{\cal T}^{-1} )_{k,k'}\, <\chi_{k'}| \, ,
  \end{equation}
   where $   {\bf \Delta }_{\cal T}$ is the overlap matrix of the test functions
  viz., $({ {\bf \Delta} }_{\cal T})_{k, k'} \, =\,   <\chi_k|\chi_{k'}>\, $.
  The overlap matrix of the two bases is denoted by $\bf \Lambda$, viz.,
  \begin{eqnarray}
  ( {\bf \Lambda})_{k, k'} \, \equiv\,   <\chi_k|\varphi_{k'}>\, .
  \end{eqnarray}
   The compatibility condition for the subspaces ${\cal S_A} \, $ and  $ {\cal S_T}\, $
   is that  ${\bf {\Lambda} } $ be non-singular.

    It will be convenient
     to introduce  a primitive basis $\{\xi_k(q,x)\, , k=1,2,...,K\, \}$,  and
     to  generate various choices of expansion and test functions
     by transforming the primitive basis
     under the action of appropriate operators (like $V$ and $G_0$)
     from the scattering theory.
     This idea is  similar  in spirit to  the concept of  {\it Riesz Bases}
     used in the theory of  approximations [26].
      Denoting with ${\cal S_{\pi}}$  the subspace spanned
     by the primitive basis $\{\xi_k(q,x) \}$,
      possibilities for the approximation  space
    include  choices like
   ${\cal S_A}\, = \, U {\cal S_{\pi}}$ where  $U$ can be taken as
     $I$, or $V$,  or even $VG_0$, depending on the nature of the
    primitive basis.
   Similarly, the test space can be generated via
   ${\cal S_T}= \, U'{\cal S_{\pi}}\, $,
   where  possible choices for $U'$ include
   operators $I$ and $V$.

  We next define an operator $\Gamma$ by
  \begin{equation}
  {\Gamma}\, =\, \Sigma_{k=1}^{K}\,  \Sigma_{k'=1}^{K}\,
   |\varphi_k> ({ \bf \Lambda}^{-1})_{k,k'}\, <\chi_{k'}| \, .
  \end{equation}
    Note
   that $\, \Gamma |\varphi_k> \, =\, |\varphi_k>\ $, and
   $\ \  <\chi_k|\Gamma\, = \,<\chi_k|\ $.  Since  $\Gamma$ has
   the idempotency property $\ \Gamma^2=\Gamma \ $, it is a projector.
   However,  $\ \Gamma \,  \neq \, \Gamma^{\dagger} \ $,
   in general. Such projection operators are referred
    to as {\em oblique} projectors [21].
   Oblique projections have received  less
   attention in the past than the orthogonal projections,
   but are recently becoming an important tool
   in, e.g.,  numerical linear algebra
   [21, 22]
    and signal processing [23].

 The  weighted-residual approach to the LS2D equation seeks an approximate
 solution $T^{WR}$ in ${\cal S_A}$ :
  \begin{equation}
 <qx|\, T^{WR}\, | q_0x_0> \, =\, \Sigma_k\, < qx|\varphi_k>\,  c_{k}(q_0,x_0)\ ,
  \end{equation}
  where $c_{k}(q_0,x_0) \, $ are the unknown expansion coefficients.
  In operator form,
   $ T^{WR}\,=\, {\cal P_A}T\, $.
   Use of $ T^{WR}\, $
   in the LS equation $T-VG_0T-V\, =\, 0$ gives rise to
   a non-zero residual (or error) function $\varepsilon (q,x)$:
  \begin{equation}
  \varepsilon (q,x)\,  =\,
   \Sigma_n \Sigma_m \ <qx|\,  (I-VG_0)\, |\varphi_k> \ c_{k}(q_0,x_0)\,
    - \, <qx|V|q_0x_0>\, .
  \end{equation}
  In weighted residual methods,
  one demands that  the residual error function
  $\varepsilon (q,x)$
    be orthogonal to the space of
  test functions:
   \begin {equation}
   <\chi_k|\varepsilon> \, =\, 0\, ,\ \ k=1,2,...,K \, .
   \end{equation}
    In operator form, this requirement  corresponds to
    \begin{displaymath}
     \, P_T \,(P_A T-V-VG_0P_A T)\, =0\, .
     \end{displaymath}
    This leads
   the following system  of linear equations for the expansion coefficients
   \begin{displaymath}
   \Sigma_{k'}  \, ({\bf D}^{-1})_{k, k'}\, c_{k'}(q_0,x_0)\, = \,
   <\chi_{k}|V|q_0x_0>\,
   \end{displaymath}
  where
  \begin{equation}
    ({\bf D}^{-1})_{k, k'}\, =\, <\chi_{k}|I-VG_0|\varphi_{k'}> .
\end{equation}
  Upon solving for expansion coefficients $\{c_{k} \}$ and using them in Eq.(7), we obtain
  \begin{equation}
  <qx|T^{WR}|q_0x_0> = \, \Sigma_k   \, \Sigma_{k'}
  \, <qx|\varphi_{k}>\, {\bf D}_{k, k'}\,
   <\chi_{k'}|V|q_0x_0>\, ,
   \end{equation}
   which represents a separable expansion of rank $K$. We can easily verify that $T^{WR}$
    is the exact $T$-operator for the approximate
    ( left-projected)potential $V^{L}\,=\, \Gamma V\, $, which
    reads in explicit notation
    \begin {equation}
    <qx|V^{L}|q'x'>\, =\, \sum_k\, \sum_{k'}\
    <qx|\varphi_k> ({\bf \Lambda}^{-1})_{kk'}\, <\chi_{k'}|V|q'x'> \ .
    \end{equation}
     Note that, $<\chi_k|V^{L}\, =\, <\chi_k|V \, $ . That is, $V^L$ is approximate
      only as far as  the left off-shell
      behaviour of $V\, $ is concerned.

      We note in passing that this procedure could
      be carried out in exactly the same manner
    for the three dimensional LS equation, Eq. (2).
     Taking  the approximation and test bases
    as functions of the full momentum vector
    (i.e., $\varphi_k(q,x,\phi)$ and $\chi_k(q,x,\phi)$, we would obtain
      \begin{equation}
  <q x \phi |T^{WR}|q' x' \phi'> = \, \Sigma_k   \, \Sigma_{k'}
  \, <qx\phi|\varphi_{k}>\, {\bf D}_{k, k'}\,
   <\chi_{k'}|V|q'x'\phi'>\, ,
   \end{equation}
    where the matrix elements involving  $T, V$ and $G_0$
    are now to be understood in
     the original three-dimensional sense.
     Although in this article we will not pursue
     this version of the method
       any further, it is conceivable that this separable form
      might provide a convenient way to generate
      the fully three-dimensional  $T$-matrix elements
      $\, <{\bf q}|T(E)|{\bf q}'> \, $ (rather than the reduced elements)
      as needed  in the context of three-particle calculations
       employing Jacobi momenta vectors directly [2,13].

   \begin {center}
    {\bf IV. GALERKIN, COLLOCATION and SCHWINGER-VARIATIONAL METHODS }
   \end{center}

    There are a multitude of possibilities for choosing the expansion and
    test bases of the weighted residual method. In this article,
    we will explore and computationally test
    only a few of these possibilities.
     The primitive basis  in our work is taken as
     local low-order (in fact, quadratic)
     piecewise polynomials (of the type used,  e.g., in  finite element methods)[18,19].
    The (orthogonal) projector onto the subspace
     ${\cal S}_{\pi}$ spanned by the primitive basis
    is given as
    \begin{equation}
  {\cal P}_{\pi}\, =
  \, \Sigma_{k=1}^{K}\,  \Sigma_{k'=1}^{K}\,
  |\xi_k>\,  ({\bf \Delta}_{\pi}^{-1})_{k,k'}\, <\xi_{k'}| \, .
  \end{equation}
  Here ${\bf \Delta}_{\pi}$ is the overlap matrix of the primitive basis, i.e.,
   $({\bf \Delta}_{\pi})_{kk'}\, =\, <\xi_k|\xi_{k'}>\, $.

   The  Galerkin method follows from taking both expansion and test bases
   as the primitive basis:
   $\varphi_k (q,x)\,
   =\, \xi_k (q,x) \, $ and $\chi_k (q,x)\,  =\, \xi_k (q,x)\, $.
    The Galerkin approximation $T^G$ for $T$ reads
   \begin {equation}
   <qx|T^{G}|q_0x_0>\,  = \, \Sigma_k   \, \Sigma_{k'}
  \, <qx|\xi_{k}>\, ({\bf D}^{G}) _{k, k'}\,
   <\xi_{k'}|V|q_0x_0>\, ,
   \end{equation}
   where
   \begin{displaymath}
   [({\bf D}^{G} )^{-1}]_{k, k'}\, =\, <\xi_{k}|1-VG_0|\xi_{k'}> .
   \end{displaymath}
    We note that
    $T^{G} $ is the exact T-operator
    for the projected potential $V^{G}\, \equiv\, {\cal P}_{\pi}V$, which
    may be termed as the   left-sided projection of the operator $V$.
    We note in passing that another version of the Galerkin method follows
    from the right-sided projection $V{\cal P}_{\pi}$.
    Yet another approximation scheme can be
     based on the two-sided projection
    ${\cal P}_{\pi}V{\cal P}_{\pi}$. (In the terminology of Ref. [25],
     this represents an {\it outer}-projection approximation.)

    For the collocation method,  we take
   $\varphi_k (q,x)\, =\, \xi_k (q,x) \, $
   and require that the error function $<qx|\varepsilon> $
   of Eq.(14) vanish on a set of $K$ collocation points
   on the computational $q$-$x$ domain.
   Let $\{q_{Cn}\, , \, n=1,2,3,...,N\, \}$ and
   $\{x_{Cm}\, , \,   m=1,2,3,...,M\, \}$ be the sets of collocation points for
   the $q$ and $x$ variables, respectively. Here $N$ and $M$ are such that $K=NM$.
   A natural choice of for these collocation points
   for a primitive basis of
   piecewise quadratic polynomials is discussed in Sec. V.
   Test  functions $\chi_k (q,x)\, $ of the collocation method are the
   delta functions located at the  collocation points, viz.,
   \begin{displaymath}
    \chi_k (q,x)\,  =\, \delta(q-q_{Cn})\, \delta(x-x_{Cm}) \, ,
    \end{displaymath}
     where index $k$ stands for $(n,m)$.
    The collocation  approximation $T^C$ for $T$ reads
   \begin {equation}
   <qx|T^{C}|q_0x_0>\,  = \, \Sigma_k   \, \Sigma_{k'}
  \, <qx|\xi_{k}>\, ({\bf D}^{C}) _{k, k'}\,
   <\chi_{k'}|V|q_0x_0>\, ,
   \end{equation}
   where
   \begin{displaymath}
   [({\bf D}^{C} )^{-1}]_{k, k'}\, =\, <\chi_k|1-VG_0|\xi_{k'}> \, .
   \end{displaymath}
   The collocation method has the advantage that
   the numerical construction of
   the matrix ${\bf D}^{C}$ is considerably easier
   than than that of  ${\bf D}^{G}$.

   The  Schwinger variational result for the T-operator
   follows from the
   choices  $\varphi_k (q,x)\,  =\,  <qx|V|\xi_k > \, $
    and $\chi_k (q,x)\,  =\, \xi_k (q,x)\, $.
   Employing these choices in Eq. (16), we obtain
   \begin {equation}
   <qx|T^{SV}|q_0x_0> = \, \Sigma_k   \, \Sigma_{k'}
  \, <qx|V|\xi_{k}>\, ({\bf D}^{SV}) _{k, k'}\,
   <\xi_{k'}|V|q_0x_0>\, ,
   \end{equation}
   where
   \begin{displaymath}
   [({\bf D}^{SV} )^{-1}]_{k, k'}\, =\, <\xi_{k}|V-VG_0V|\xi_{k'}> .
  \end{displaymath}

    It is a well known fact that $T^{SV}$ is the exact T-operator
    for the finite-rank approximate potential
   \begin{displaymath}
   V^{IP}\equiv V\, {\cal P}_{\pi} \, ({\cal P}_{\pi}  V {\cal P}_{\pi} )^{-1}\, {\cal P}_{\pi} V.
   \end{displaymath}
   Such operator approximations are known  as
   {\it inner-projection} approximations
   in the Quantum Chemistry literature [25, 26]. In explicit notation,
   \begin{equation}
   V^{IP}(q,x;q',x')\, = \, \Sigma_{k=1}^K\, \Sigma_{k'=1}^K\,
   <q x|V|\xi_k>\,({\bf V^{-1}} )_{k,k'}\, <\xi_{k'}|V|q'x'> \, .
   \end{equation}
  The inner-projection approximation
  has the interesting property that
  $V^{IP}{\cal P}_{\pi}\,  =\ V{\cal P}_{\pi}$
  and ${\cal P}_{\pi} V^{IP} \ =\ {\cal P}_{\pi} V \, $.
  That is, $V^{IP}$
  is not restricted to the approximation space ${\cal S}_{\pi}$.
  Inner projection approximation can also be viewed as
  an oblique projection.
  Defining the   oblique projector
  $\Gamma^V\, =\,\Sigma_{k=1}^K\, \Sigma_{k'=1}^K\,
   |\xi_k>\,({\bf V^{-1}} )_{k,k'}\, <\xi_{k'}|V\,  $,
   we have $V^{IP}\, =\, V\, \Gamma^V\,
   =\, {\Gamma^V}^\dagger\, V$.

  A variant of the Schwinger variational method can be obtained by taking
  $\chi_k (q,x) =\, <qx|G_0|\xi_k >$ and $\varphi_k (q,x) = <qx|VG_0|\xi_k > \, $.
  We will refer to this scheme as Schwinger Variational method
   with a $G_0$-weighted basis (in short, $SVG0$ method). The resulting expression for the
   $T$-matrix is
   \begin{equation}
    <qx|T^{SVG0}|q_0x_0>\,  = \, \Sigma_k   \, \Sigma_{k'}
  \, <qx|VG_0|\xi_{k}>\, ({\bf D}^{SVG0}) _{k, k'}\,
   <\xi_{k'}|G_0V|q_0x_0>\, ,
   \end{equation}
   where
    \begin{displaymath}
   [({\bf D}^{SVG0} )^{-1}]_{k, k'}\, =\, <\xi_{k}|G_0VG_0-G_0VG_0VG_0|\xi_{k'}> .
     \end{displaymath}

     Another choice for the expansion and test bases would be
   $\varphi_k (q,x) = <qx|V|\xi_k > \, $ and
   $\chi_k (q,x)\,  =\, \delta(q-q_{Cn})\, \delta(x-x_{Cm}) \, $, respectively.
   The ensuing
   approximation will be referred to as the {\em hybrid collocation (HC)}
   method. The expression for $T^{HC}$ reads
   \begin{displaymath}
   <qx|T^{HC}|q_0x_0> = \, \Sigma_k   \, \Sigma_{k'}
  \, <qx|V|\xi_{k}>\, ({\bf D}^{SV}) _{k, k'}\,
   <q_{Cn'}x_{Cm'}|V|q_0x_0>\, ,
   \end{displaymath}
   where
   \begin{displaymath}
    [({\bf D}^{HC} )^{-1}]_{k, k'}\, =\,
    <q_{Cn}x_{Cm}|V-VG_0V|\xi_{k'}>\, ,
  \end{displaymath}
   where $k=(n,m)$ and $k'=(n',m')$
   whenever $k$ and $k'$ enumerate the  collocation points.
   This method stands to the SV method  in the same way as collocation
   method stands to the Galerkin method.
   Construction of ${\bf D}^{HC}$ involves one
   integration (over $q$-$x$ domain)  less than the construction
   of ${\bf D}^{SV}$.

\begin {center}
{\bf V. FINITE ELEMENT BASES AND COMPUTATIONAL IMPLEMENTATION }
\end{center}

  The primitive approximation space ${\cal S}_{\pi}$ is constructed as a direct product
  space: $\, {\cal S}_{\pi}\, =\, {\cal S}_{\pi q}\otimes{\cal S}_{ \pi x}\, $. The basis for
  the $N$-dimensional
   space  ${\cal S}_{\pi q}$ is  denoted as
  $\,  \{ f_n(q)\, , \, n=1,2,...,N\, \}\,  $, whereas $\, {\cal S}_{\pi x}\, $
  is $M$-dimensional
  and spanned by
   $\, \{ h_m (x)\, ,\,  m=1,2,...,M\,  \}\, $.
   Hence, ${\cal S}_{\pi}\, $ is of dimension $K=NM$, and spanned by
   the direct-product basis  $\,  \{ \xi_{nm}(q,x) \, \equiv \, f_n(q)h_m(x)\, \}\, $.
    The basis sets in the $q$ and $x$ variables  are  linearly independent,
    but not necessarily orthonormal.   The overlap matrix  is a direct-product  matrix:
    ${\bf \Delta}_{\pi}\,=
   \, {\bf \Delta}_{\pi q}\, \otimes \, {\bf \Delta}_{\pi x}$ ,
   where
   $({\bf \Delta}_{\pi q})_{n, n'} \, =\, <f_n|f_{n'}>\, $ and
   $({\bf \Delta}_{\pi x})_{m, m'} \, =\,     <h_m|h_{m'}>\, $.
    The inner products are taken as
    $\, <f_n|f_{n'}> \,=\, \int_0^{\infty}\, q^2\, dq\, f_{n}^{\ast}(q)\, f_{n'}(q)\, $ and
    $\, <h_m|h_{m'}>\,=\, \int_{-1}^{1}\, dx\, h_{m}^{\ast}(x)\, h_{m'}(x)\,
    $.

   The primitive basis functions will be taken
   as  local  piecewise quadratic polynomials [19]
    defined over a grid, as in the finite element method.
    For our purposes, quadratic interpolates are found
      to provide sufficient flexibility,
      although higher order interpolates
    like cubic hermites or cubic splines [19] could also be used.
     Interestingly, in Ref.[27], piecewise constant functions
    (the so-called {\it hat} functions)
    over a grid has been shown to be quite efficient
    to solve partial-wave (single-variable)  scattering
    integral equations via a projection method (similar
    to the outer-projection method mentioned in the previous
    section).

     To define the $q$-grid over which piecewise polynomials are constructed,
     we divide the  domain into two intervals:
     $[0,\, 2q_0]$,  and $[2q_0, \infty )$.
     This scheme is adopted to
     treat the singularity at $q_0$ as
    symmetrically as possible, and to use a denser grid
    for $q$ in the vicinity of $q_0$.
     To this end, the first interval $[0,\, 2q_0]\, $,
     is   subdivided into $I_1$  (equal) subintervals (finite
     elements).

     On the other hand, the  second interval
     $[2q_0,\infty)$ is mapped to $[-1,+1]$
     via the  transformation
     \begin{equation}
     u\, =\,  \frac{q-2q_0-f}{q-2q_0+f} \, , \ \  \mbox{or}\ \  q\, =\, 2q_0\,  +\, f\, \frac {1+u}{1-u}\ ,
     \end{equation}
      where $f$ is a scale factor.
       By setting an upper limit $u_{max}\, (<1)$
      to  the $u$ variable, the $q$-variable is cut off
      at some large but finite value $q_{max}$.
      Depending on the values used for  $u_{max}$
      and the scale factor $f$,
      momentum cutoff  in our calculations ran into
       several thousand atomic units.
      This scheme paves the way for a discretization of the semi-infinite interval
      $[0,\infty)$ with relatively  few finite elements.
       The interval $[-1,u_{max}]$ is divided
       into $I_2$ equal finite elements (which, however,
      correspond to
      a non-uniform partitioning in the $q$-variable.)

     Let $\{ Q_1,\,  Q_2,\,  ... ,\,Q_{I_{1}+1}   \}$ be the set of break points for  a
   partition of
   the  interval $[0,2q_0]$ into $I_{1}$ finite elements.
   Here $ Q_{1}\, =\, 0\, $ and
   $\, Q_{I_{1}+1}\, =\,2q_{0}\,$. The midpoint of the
   $i$th finite element $[Q_{i},Q_{i+1}]$ is
   denoted  ${Q}_{i+1/2}\ \, ,  i=1,2,..., I_{1}\, $.
   For the second interval $[2q_0,q_{max}]$,
    let  $\{ u_1,\,  u_2,\,  ... ,\,u_{I_2+1}   \}$  be the break points for
    a partition of
   the   corresponding interval
   $[-1,u_{max}]$ of the transformed variable $u$.
    Here $u_1=-1$, and $u_{I_2+1}=u_{max}$.
   The breakpoints $\{u_{i'} \}$ and mid-points $\{ {u}_{i'+1/2} \}\, $,
   $i'=1,2,...,I_2\,$,
   of this partition
    are  mapped via Eq. (26), respectively,  to $\{Q_i\}$ and $\{Q_{i+1/2)} \}$, where
   $i=i'+I_{1}\,$.

    The total number of finite elements covering the computational interval
    $[0,q_{max}]$ is $I \ (\equiv I_1+I_2) $. For the calculations reported in the next section,
    the choice $I_2=3I_1$ (hence $I=4I_1)$ was found adequate after
    some experimentation.
    Collecting and ordering
    the break-points and mid-points of all the  finite elements together,
    we form the set
    $\, \{{q_{1},q_{2},..., q_{N}}\,  \} \, $ of
    grid points,
     where
    $\, N=2I+1\, $,  $\, q_{N}\, =\, q_{max} \, $, and
   $\, q_{2i-1}\, =\, Q_{i}\, , \,   q_{2i}\,
    =\, { Q}_{i+1/2}\, $, for $\, i=1,...,I \, $.
     This set ( to be referred to as the {\it grid})
     provides the setting
     for the definition of the $q$-basis $\{ f_n \} $.
     Each basis function $f_n(q)$ will be
     centered at its corresponding  grid point $q_n$
      and will satify the cardinal property
      $f_n(q_m)\, =\, \delta_{nm}\, , \ n,m=1,2,...,N\, $.
      This set of grid points
        also provide a natural choice
        as the collocation points $q_{Cn}$ for the $q$-variable.

      The basis functions associated with the first $2I_{1}+1$
      grid points are taken as  piecewise quadratic  functions of $q$,
      while the ones associated with the grid points in the
      in the interval $[2q_0,q_{max}]$ are
      taken as piecewise quadratic polynomials
      in the transformed variable $u$.
      These local piecewise polynomials are
      best described in terms of a local variable $s$, defined separately
      for each finite element. For the
      finite elements covering $[0,2q_0]$,
      the finite-element interval $[Q_{i},Q_{i+1}]$
      is mapped to $[-1,1]$
       via $s=(2q-Q_{i}-Q_{i+1})/(Q_{i+1}-Q_{i})\,$. For finite
       elements in $[2q_0,q_{max}]$, we map the $u$-variable finite element
       $[u_{i},u_{i+1}]$
       into $[-1,+1]$ via the map $s=(2u-u_{i}-u_{i+1})/(u_{i+1}-u_{i})\,$.

     In terms of  the  local variable $s$,
     the basis functions associated with the breakpoints read
     \begin {eqnarray}
     f_{2i-1}(q)\, =\ \left \{
     \begin{array}{cr}
     -s(1-s)/2 & \mbox{for}\ \ \ Q_{i} \, < \, q \,< \, Q_{i+1} \\
     s(1+s)/2  & \mbox{for}\ \ \ Q_{i-1} \, < \, q \,< \, Q_{i} \\
     0  &  \mbox{otherwise} \ \, \ \
  \end{array}
  \right.
   \end{eqnarray}
\noindent for $i=1,2,...,I+1\, $; while the functions associated
with the midpoints of finite elements  have the form
\begin {eqnarray}
     f_{2i}(q)\, =\ \left \{
     \begin{array}{cr}
     1-s^2 & \mbox{for}\ \ \ Q_{i} \, < \, q \,< \, Q_{i+1} \\
     0  &  \mbox{otherwise} \ \, ,
  \end{array}
  \right.
   \end{eqnarray}
\noindent for $i=1,2,...,I\, $. These functions are depicted, e.g., in Ref. [28]
 where they have been used to discretize the momentum space in the context
 of a time-dependent wave-packet calculation of partial-wave S-matrix elements.

    We note that each basis function has a finite support:
     two finite elements  for functions associated with breakpoints, and
     one finite element for functions corresponding to the midpoints.
   A characteristic (cardinal) property of these finite-element basis functions is that
     $f_n(q)$  vanishes  at all grid points  except at $q=q_{n}\,$
     where it has the value of unity:
      $\, f_n(q_{n'})\, =\,\delta _{nn'}\, $.

       The discretization of the $x$-variable  proceeds similarly
       to that of the $q$-variable.
       The interval $[-1,1]$ is
       partitioned into $J$ subintervals (elements)
           by specifying breakpoints  $\{ {X_{1},X_{2},..., X_{J+1}}\,  \}\, .$
        Here, $\,X_{1}\, =\, -1\, $ and $\, X_{J+1}\, =\, 1 \, $.
        In contrast to $q$-variable, the placement of the breakpoints for the $x$-partition is
          uniform. (A non-uniform x-grid  is of course possible and may be more appropriate
          in some cases.)\
          The midpoint of the
      $i$th interval $[X_{i},X_{i+1}]$ is
   denoted  ${ X}_{i+1/2}$. Collecting
   the breakpoints and midpoints together, we  define
    the set of grid points
   $\, \{ {x_{1},x_{2},..., x_{M}}\,  \}\, ,$ where
   $\, M\, =\, 2J+1\, $,  $\ x_{M}=\, 1 \, $, with
   $\, x_{2i-1}=X_{i}\, $ and  $ \ x_{2i}\,  =\, { X}_{i+1/2}\, $
    for  $\, i=1,...,J \, $.
   The $x$-basis $\, \{ h_m(x)\,  \} \, $
    consists of $M$ piecewise quadratic  functions defined on this grid.
    Again, there is one quadratic function
     associated with each grid point and, in terms of
      the local variable $s$, defined by
      $s=(2x-X_{i}-X_{i+1})/(X_{i+1}-X_{i})\, $,
     they have  exactly the same functional form as in Eqs. (26) and (27) (with of course
      $q$ and $Q$ replaced by $x$ and $X$, respectively) .

     Of course, other basis functions than piecewise
     quadratic polynomials could be used
     for the variables $q$ and $x$.
     Use of global bases (like gaussians on the grid, or sinc functions)
     in place of
     localized bases for $q$ and/or $x$ is a possibility.
      For instance, the use of Legendre polynomials for $x$ would
     be equivalent to the partial wave expansion.
     Another possibility is to employ
      more sophisticated finite-element bases,
      such as the higher order piecewise polynomials
      adopted to    more complicated grids (such as decomposition into triangles) of  the
       computational domain  on the $q-x$ plane.
      For example,  Ref. [3] used fifth-degree polynomials
      in radial coordinate $r$ and polar angle $\theta$ over a rectangular grid
      on the $r-\theta$ plane
      to solve the two-dimensional Schrodinger equation
      in coordinate space.

     Collision energies used in   our calculations correspond to
      the  on-shell momentum $q_0$  having values $0.5\, $
     and $2.0\,$.  For each  finite-element
    $\, [\, Q_{i},Q_{i+1}\, ] \, $,   a set of $n_q$ Gauss-Legendre quadrature points are chosen
    by transforming  to the local variable
   $s$ defined earlier.
   The Gauss-Legendre quadrature points   for all elements
   are then combined and ordered to form a composite quadrature
   rule with
   the set of quadrature points
   $\, \{ \, q_{\alpha}  , \,  \alpha\, = \, 1,2,...,N_q\, \} \, $,
   where
    $N_q\, = I\,n_q $.
    The quadrature weights are similarly collected in the set
    $\, \{ w_{\alpha}  , \,  \alpha\, =\, 1,2,...,N_q\, \}\, $.
    In the calculations reported in the next section,
    $n_q$ was typically taken as 8.
    For the $x$-variable, in each finite element
     $\, [\, X_{j},X_{j+1}\, ] \, $, we take $n_x $
    Gauss-Legendre quadrature points.
    The quadrature points and their weights are collected, respectively,
     in the sets
    $\{ x_{\beta}\, , \,  \beta\, =\, 1,2,...,N_x\, \}  $, and
    $\{ \rho_{\beta}\, , \, \beta \, =\, 1,2,...,N_x\, \} $,
    where $N_x=J\,n_x\,$. In our calculations, typically $n_x=8$ (and $N_x=80$)
    was sufficient to obtain 6 digit accuracy.

     Singular integrals involved in matrix elements like
     $<\xi_{k}|VG_0V|\xi_{k'}>$,
     $<\xi_{k}|G_0VG_0|\xi_{k'}>$, and $<\xi_{k}|G_0VG_0VG_0|\xi_{k'}>$ are handled
     by the well-known subtraction technique. For instance, the matrix element
     $<f_nh_m|VG_0V|f_{n'}h_{m'}>$ is first written as
     \begin{eqnarray}
      <f_nh_m|VG_0V|f_{n'}h_{m'}>\,= \,2\mu \ {A}_{nm,n'm'} \,
       -\, i\pi \mu q_0 \, B_{nm,n'm'}(q_0)  \nonumber
      \end{eqnarray}
      where
      \begin{eqnarray}
       A_{nm,n'm'} \, & =   & \, {\cal P}\int_0^{q_{max}} \,  dq
       \, \frac {\,q^2\, B_{nm,n'm'}(q)} { q_0^2-q^2 \, } \nonumber \\
       B_{nm,n'm'}(q)\,&  =\, &
      \int_{-1}^{1}\, dx\,
       <f_nh_m|V|qx>\,   <qx|V|f_{n'}h_{m'}>\, \nonumber
      \end{eqnarray}
      where ${\cal P}$ denotes principle-value integral.
      The matrix element  $A_{nm,n'm'} $ is then
      rewritten as the sum of  non-singular and singular parts:
      \begin{eqnarray}
              A_{nm,n'm'} & =  & A^{(ns)}_{nm,n'm'}\, +\, A^{(s)}_{nm,n'm'}\ , \nonumber
      \end{eqnarray}
      where
      \begin{eqnarray}
      A^{(ns)}_{nm,n'm'}  & =  & \int_0^{q_{max}}\,  dq\
      \frac { q^2 \,  B_{nm,n'm'}(q)\,   -\, q_0^2 \,  B_{nm,n'm'}(q_0) }
       { q_0^2-q^2  }\, ,  \nonumber \\
       A^{(s)}_{nm,n'm'}  & =  & B_{nm,n'm'}(q_0)\,
       \int_0^{q_{max}}\, dq\  \frac{q_0^2}{ q_0^2-q^2} \,
        =\, B_{nm,n'm'}(q_0) \,\frac{q_0}{2} \ln \frac{q_{max}+q_0}{q_{max}-q_0}\ . \nonumber
     \end{eqnarray}
      The integrals involved  in $B_{nm,n'm'}(q)$ and $A^{(ns)}_{nm,n'm'} \,$  are
      approximated by quadrature:
     \begin{eqnarray}
     B_{nm,n'm'}(q)   & \approx   & \Sigma_{\beta=1}^{N_x}\, \rho_{\beta}\,
      <f_nh_m|V|q x_{\beta}>\, <q x_{\beta}|V|f_{n'}h_{m'}>\,
      \nonumber \\
      A_{nm,n'm'} & \approx   & \Sigma_{\alpha=1}^{N_q}\, \, w_{\alpha}\,
       q_{\alpha}^2\,
      \frac {B_{nm,n'm'}(q_{\alpha} }
      { q_0^2-q_{\alpha}^2 \, } \, +\, C_{sing}\, q_0^2\, B_{nm,n'm'}(q_0) \nonumber
      \end{eqnarray}
      \noindent  where
      $C_{sing}$ represents the difference between exact and quadrature evaluations of the
      singular integral  ${\cal P}\int_{0}^{q_{max}} dq / (q_0^2-q^2)\, $:
      \begin{displaymath}
      C_{sing}\,= \,\,\frac{1}{2q_0} \ln {\frac{q_{max}+q_0}{q_{max}-q_0}} \, -
       \, \sum_{\alpha=1}^{N_q}\, \frac{w_{\alpha}}{q_0^2-q_{\alpha}^2}
      \end{displaymath}

      The reference results against which the results of
      weighted-residual  methods will be compared
   are  obtained by solving the two-variable integral equation
  via the Nystrom method (i.e., the quadrature discretization
  method). To prepare for the quadrature discretization,
  we define
  \begin{equation}
    {\cal K}(q,x;q'x'|q_0) \ =\
  q'^2\,V(q,x;q',x')\, -\, q_0^2\, V(q,x; q_0,x')
      \end{equation}
   and rewrite Eq. (7) as
  \begin{eqnarray}
  T(q,x;q_0,x_0)\,&  = & \, V(q,x;q_0,x_0)\,  \nonumber \\
    &+\,   & \, 2\mu\,  \int_{0}^{q_{max}}\,  dq'\,  \int_{-1}^{1}\,  dx'
  \ \frac {{\cal K}(q,x;q'x'|q_0) } {q_0^2-q'^2 }\ T(q',x';q_0,x_0)  \nonumber \\
  & +\,  &  2\mu q_0^2 \, \int_{-1}^{1}\,  dx' V(q,x; q_0,x') T(q_0,x';q_0,x_0)
  \,\int_{0}^{q_{max}}\,  dq'\,
   \frac{1}{q_0^2-{{q'}^2}+i0}\  , \nonumber \\
        &    &
  \end{eqnarray}
  \noindent where a term involving
  the singular integral
  $\int_{0}^{q_{max}}\,  dq'\,({q_0^2-{q'}^2+i0})^{-1}$ has been added and subtracted.
   The first term on the right hand side  with the subtracted kernel is now non-singular and
   can be approximated by quadrature. The second integral is to be evaluated analytically.
   This scheme has been tested and verified  against  other subtraction methods.
   For example,  a three-dimensional generalization
   of Kowalski-Noyes method $[16]$ has also been used
   and will be described elsewhere.

  The same set of quadrature points used in the
  implementation of the weighted-residual methods  are used to discretize Eq. (30).
   Replacing the integral by the quadrature sum and
  collocating at  $q'=q_\alpha$, $x'=x_\beta$, and $q'=q_0$, we obtain  a set of
  $(N_q+1)N_x$ linear equations for
  $T(q_\alpha,x_\beta;q_0,x_0)$ and $T(q_0,x_\beta;q_0,x_0)$ as
  \begin{eqnarray}
   T(q_{\alpha},x_{\beta};q_0,x_0)\, & =
   & \, V(q_{\alpha},x_{\beta};q_0,x_0)\,  \nonumber  \\
      & + &  2\mu\,
      \Sigma^{N_q}_{ {\alpha'}=1 } \,
      \Sigma^{N_x}_{ {\beta'}=1}
      \, {q_{\alpha'}}^2 {w_{\alpha'}}\,
       {\rho_{\beta'}}\,
       \frac {\, V( q_{\alpha},  x_{\beta}; q_{\alpha'},  x_{\beta'} )\,
       T(    q_{\alpha'},  x_{\beta'}; q_{0},  x_{0} )\,}
       { { q_0}^2\, - \, {q_{\alpha'}}^2\, }  \nonumber  \\
        & +  & \, C_{pole} \,
       \Sigma^{N_x}_{ {\beta'}=1}
       {\rho_{\beta'}}\,
       \, V( q_{\alpha},  x_{\beta}; q_{0},  x_{\beta'} )\,
       T(    q_{0},  x_{\beta'}; q_{0},  x_{0} )\, ,
    \end{eqnarray}
    \noindent and
    \begin{eqnarray}
   T(q_{0},x_{\beta};q_0,x_0)\, & =
   & \, V(q_{0},x_{\beta};q_0,x_0)\,  \nonumber  \\
      & + &  2\mu\,
      \Sigma^{N_q}_{ {\alpha'}=1 } \,
      \Sigma^{N_x}_{ {\beta'}=1}
      \, {q_{\alpha'}}^2\, {w_{\alpha'}}\,
       {\rho_{\beta'}}\,
       \frac {\, V( q_{0},  x_{\beta}; q_{\alpha'},  x_{\beta'} )\,
       T(    q_{\alpha'},  x_{\beta'}; q_{0},  x_{0} )\,}
       { { q_0}^2\, - \, {q_{\alpha'}}^2\, }  \nonumber  \\
        & +  & \,C_{pole}\,
       \Sigma^{N_x}_{ {\beta'}=1}
       {\rho_{\beta'}}\,
       \, V( q_{0},  x_{\beta}; q_{0},  x_{\beta'} )\,
       T(    q_{0},  x_{\beta'}; q_{0},  x_{0} )\, .
    \end{eqnarray}
    \noindent where $C_{pole}\, =\,
     \,2\mu q_0^2\,  C_{sing}\, -\,  i\pi \mu \, q_0\, .$

    Once $T(q_\alpha,x_\beta;q_0,x_0)$ and $T(q_0,x_\beta;q_0,x_0)$
    are obtained by solving the above set of
     linear  equations, the matrix elements $T(q, x; q_{0},  x_{0} )$
     for arbitrary values of $q$
     and $x$ can now be obtained from
     \begin{eqnarray}
     T(q,x;q_0,x_0)\, & =
   & \, V(q,x;q_0,x_0)\,  \nonumber  \\
      & + &  2\mu\,
      \Sigma^{N_q}_{ {\alpha}=1 } \,
      \Sigma^{N_x}_{ {\beta}=1}
      \, {q_{\alpha}}^2\, {w_{\alpha}}\,
       {\rho_{\beta}}\,
       \frac {\, V( q,  x; q_{\alpha},  x_{\beta} )\,
       T(    q_{\alpha},  x_{\beta}; q_{0},  x_{0} )\,}
       { { q_0}^2\, - \, {q_{\alpha}}^2\, }  \nonumber  \\
        & +  &   C_{pole}\,
       \Sigma^{N_x}_{ {\beta}=1}
       {\rho_{\beta}}\,
       \, V( q,  x; q_{0},  x_{\beta} )\,
       T(    q_{0},  x_{\beta}; q_{0},  x_{0} )\, .
    \end{eqnarray}

      \begin {center} {\bf VI. RESULTS  } \end{center}

      For our calculations, we use  the Hartree  potential

      \begin{displaymath}
        V(r)\, =\, V_0 \, e^{-\lambda r}\, \left(\, 1 +\, \frac{1}{r} \right)\ .
     \end{displaymath}
     The values used for the potential parameters are
     $V_0=-2.0$ and $ \lambda =-2.0$,
     and  the  reduced mass is $\mu=0.5$.\\

     The  momentum-space  representation of this potential is given as
     \begin{displaymath}
        V({\bf q},{\bf q}')\,   =\,
        \frac{\lambda V_0 }{\pi^{2}}\,  \frac{1} { [ \, ( {\bf q}-{\bf q}')^2+\lambda^2\, ]^2 }
        \,  - \frac{V_0}{2\pi^{2}}\, \frac{1} {({\bf q}-{\bf q}')^2 +\lambda^2 }
     \end{displaymath}
     For this potential the azimuthal integration in Eq. (4) can be carried out analytically to give
       \begin{eqnarray}
        V(q,x;q',x')\,  & = & \,    \frac{2\lambda V_0}{ \pi}\,\,
        \frac{(q^2+q'^2-2qq'xx'+\lambda^2) }
        { [\, (q^2+q'^2-2qq'xx'+\lambda^2)^2  - 4 q^2q'^2(1-x^2)(1-x'^2)\, ]^{3/2} }  \nonumber \\
        \,& \ & - \,  \frac{V_0}{\pi} \, \frac{1}{[\,(q^2+q'^2-2qq'xx'+\lambda^2)^2\,
         -\,  4 q^2q'^2(1-x^2)(1-x'^2) \, ]^{1/2} }   \nonumber
  \end{eqnarray}
  The availability of analytical form for $V(q,x;q',x')$ is not crucial. The reduced potential
  $V(q,x;q',x')$ can be generated numerically by applying an appropriate quadrature rule to the
  $\phi$ integral. In fact, for the present model,
  a composite 64-point Gauss-Legendre rule for $\phi-$integral produces
  results that are indistinguishable within 7-8 digits from those of
  the analytical reduced potential.

  Table I shows results of  Nystrom  calculations at E=0.25 and 4.0
  for two  values of  momentum cutoff $q_{max}$.
  Shown are the real and imaginary parts of the scattering
   amplitude
   \begin{displaymath}
   A(x;E)\, \equiv\,-4\pi^2\mu\, T(q_0,x;q_0, x_0=1.0;E)
   \end{displaymath}
  for three values of $x$.
  Also shown is  the average of the scattering amplitude over $x$
  (which is the s-wave component of the scattering amplitude).

  The computational parameters for the Nystrom calculations
   with $ q_{max}=30$  are as follows:
   the computational $q$-interval
   $[0,30]$  was partitioned into 20 finite elements and
    a composite quadrature rule
  constructed by taking 8 quadrature points
  per finite element.
  For the $x$-variable, the interval $[-1,+1]$
   divided into 10 finite elements and 8
   quadrature points were used per finite element.
   Thus, 160 points have been used to discretize the $q$-variable over
   the computational interval $[0,30]$, and 80 quadrature points for
   the $x$-variable.  The
   order of the coefficient matrix of the Nystrom method is
   12880. Such  systems of equations have been solved  by
   a direct out-of-core equation solver (described earlier in [28]).
   For even larger dimensions, Pade re-summation of the Born series
   generated from Eq. (29) turns out to be very efficient.
   Direct and Pade solutions agree to 8 digits.
   The results are also converged to at least 6 digits with respect
   to further variations
   of the computational parameters like number of quadrature
   points and their distribution.

   The case indicated as $q_{max}>1000$ in Table I involve mapping of
   the interval $[30, \infty]$ by the transformation (27) to $[-1,+1]$.
   Momentum cutoff is introduced by truncating $[-1,+1]$ as
   $[-1,u_{max}]$, with $u_{max}$ taken typically as 0.99.
   With scale factor $f=30$, this gives $q_{max}\approx 6000$.
   In calculations of Table I, the interval $[-1,0.99]$
   was divided into $10$ finite-elements (i.e., $I_2=10$)
   and an 8-point quadrature
   used over each finite element. That is,
   80 additional quadrature points
   have been used for the interval $[30,q_{max}]$.
   Thus the total number of quadrature points
   for the full computational $q$-interval $[0,q_{max}]$
   is 240. With 80 quadrature points used to discretize the
   $x$ integral, the number
   of equations to be solved comes out as 19280.
    The results obtained with either the out-of-core direct solver
    or Pade resummation are
    converged at least within the number of digits shown in the table
   (or better) to further variations of computational parameters.

    Table I demonstrates that results accurate to
    within 3 or 4 digits can be obtained
   if $q$-integration is cut off at 30 atomic units.
   To obtain results stable at 6 digit level
   one needs to extend the cut off beyond 1000 atomic units.
    If one attempts to discretize the
    the interval $[30,q_{max}]$ directly in the variable $q$,
    this could  lead to enormous number
    of quadrature  points and
    to an intractable computational task.
    Fortunately, however, the mapping of Eq. (25)
     makes this task tractable with relatively
     few finite elements and quadrature points.
     Similar transformations are
     applied, e.g., in Refs. [1],[5],and
     [8], to map the full interval $[0,\infty]$ to $[-1,+1]$
      or $[0,1]$. However,
      such maps do not  treat the singularity
     at $q=q_0$ with sufficient care. In our scheme,
      we separate out the low-momentum region
      for direct and careful treatment,
      and apply the mapping to the high-momentum region.

   Tables II and III shows the results of SVM calculations at $E=0.25$ and
   $E=4.0$ for various basis sizes. In these tables,
   $N$ and $M$ are the number of
   basis piecewise quadratic polynomials
   for the $q$ and $x$ variables, respectively.
   The orders of the linear equation
   systems that result from  SVM
    are significantly smaller than that of the Nystrom method.
    The dimension of    the ${\bf D}^{SV}$ matrix of Eq. (21) ranges from 99 to 861
   for calculations reported in Table II, and from 357 to 1271 in Table III.
   Also shown on these tables are the results
   of Shertzer and Temkin who have solved
   the  two-dimensional Schrodinger equation
   for the same potential
   with the finite element approach [3].
   The agreement between their  results   and ours
   is excellent.  Their results, however,  are reported
   to within 3 digits after the decimal point.
   Our Nystrom  results are stable to within at least
   6 digits after the decimal point
   to further variations of all computational parameters.

       Table IV shows the results obtained with the Galerkin method
       at $E=0.25$ using piecewise quadratic polynomials as the
       basis. Comparison of these results
       with those of Table II demonstrates that SVM exhibits better
       convergence than the Galerkin method. To obtain 6 digit accuracy,
       Galerkin method requires a  finer partitioning
       of the computational domain.
       Table V shows that results obtained with the  collocation method
       are very similar to those of the Galerkin method.

      Table VI gives the results of calculations
      employing the Schwinger Variational method
      with a $G_0$-weighted basis based on Eq. (23).
      These are to be compared with those of Table II.
      The transformation of
      the primitive $q$-basis under $G_0$ is especially effective for small bases.
      For instance, at the  $N=19$ level,
      the quality of the results  of the weighted basis
      are superior to those of the (primitive)
      piecewise quadratic basis. However, the additional $G_0$
      factors in various matrix elements  make this approach
       numerically more involved.

       Table VII shows the results obtained with the
       hybrid-collocation method.  Quality of results are very
       similar to that of SVM (listed in Table II). As collocation
        methods require less numerical integration, computational
        savings could make this hybrid method competitive among
        the various schemes considered.

        Finally, Table VIII shows results obtained with SVM
        using a primitive  basis of Gaussian functions for the $q$-variable. The basis
        functions $f_n(q)$ are taken as Gaussian functions
        centered on the collocation points $q_{Cn}$:
        \begin{displaymath}
        f_n(q)\, =\, e^{-a_n (q-q_{Cn})^2 }
        \end{displaymath}
         The width parameters $a_n$ were adjusted so that the effective
         support of the Gaussians extended over only a few finite
         elements around $q_{Cn}$. The choice $a_n=3/(q_{Cn+1}-q_{Cn})$
         was used in calculations reported in Table VIII.
          It is interesting to see that
         convergence pattern for this Gaussian basis
         is nearly the same as in the case of
         piecewise quadratics.

    \begin {center} {\bf VII. DISCUSSION and CONCLUSIONS  } \end{center}

    We have shown that  versatile computational schemes can be constructed
    via the the weighted-residual approach
    to directly calculate the full three-dimensional
    momentum representation of the two-particle transition operator
    without invoking angular momentum decomposition.
    With these methods, the
    accuracy of the Nystrom method with a fine
    quadrature mesh can be reached with relatively small bases,
    reducing the order of equations to be solved
    by at least an order of magnitude.
    We have demonstrated that SVM converges faster than the Galerkin
    and collocation methods.
     Of the various versions considered,
     Schwinger-variational and hybrid-collocation methods
      appear more promising. Especially, the hybrid collocation
      method combines the advantages of
      Schwinger-variational and collocation methods.
     The separable form of the $T$-operator and
     the relative ease with which arbitrary
    off-shell $T$-matrix
    elements can be generated  should make such methods
    quite attractive for use in direct momentum-vector
     approach to three-particle Faddeev equations
      without employing partial-wave decomposition [13,14].

     We have discussed the generation of expansion and test bases
     via
    transformation of  a primitive basis under
    some (invertible) operator.
     Such bases can be termed as the Riesz bases,
    following the
     terminology of  the frame theory [26].
    We have shown that employment of $\{G_0|f_nh_m> \}$
    as the basis in SVM leads to improved convergence. However,
    the appearance of additional $G0$ factors in various matrix
    elements means more numerical work to evaluate them.
    The use of other Riesz-like bases,  such as $\{G_0V |f_nh_m> \}$,
    as  expansion and/or test functions
    in the general weighted-residual
    expression (17) for $T^{WR}$ might be explored.

     The use of local finite-element bases in weighted-residual methods
     is not an inherent requirement of such methods. Other localized  bases (like
     Gaussians on a grid or  sinc functions [30])
     or global functions may also be considered.
     Another possibility is to give  up the direct-product bases and
    instead use bases that entangle  $q$ and $x$ variables.

    We have discussed the SVM , Galerkin and collocation methods
    in the spirit of the weighted-residual idea (or
    Petrov-Galerkin ansatz).
    This paves the way to view these well known approaches in a new light.
    For example, the Nystrom method can itself  be viewed as
    a collocation-type weighted-residual method.
    In our implementation of Nystrom and various weighted-residual methods
        the same quadrature scheme has been
    used to evaluate the integrals involved. With this caveat in mind,
     the (smaller) systems of linear equations
    that result from the weighted-residual methods
     can be viewed as contractions
    (or projections) of the (larger) set of equations
    of the Nystrom method. In effect, the  set  of equations
    over the $N_qM_x$-dimensional
     vector space (stemming from the Nystrom discretization)
     is  replaced
     by an  (approximate) smaller  set of equations
     on a  subspace of dimension
     $NM$. For example, one can show  that the contraction
     from Nystrom to Galerkin is affected  by the
     $(N_qM_x\times NM$)-dimensional direct-product matrix $\bf U$,
     defined as
     $ {\bf U}_{nm, \alpha\beta}\ =\ \xi_k(q_{\alpha}, x_{\beta})\, w_{\alpha}\rho_{\beta}\,
     =\, f_n(q_{\alpha})h_m (x_{\beta})\,w_{\alpha}\rho_{\beta}\,  $.
     This is similar, e.g., to the well-known connection
     between (orthogonal) collocation
     and Galerkin methods [19].
     In fact, this type of contraction is common place in numerical linear algebra.
     Although the weighted-residual idea is usually employed in the context of differential and
    integral equations, its use in numerical linear algebra leads to fruitful results as well [22].
    For instance, Krylov subspace methods [22,31] for linear systems use
      the weighted-residual idea to replace
    an original large matrix problem by a smaller approximate
    system of equations.

    For the present model,
    the number of partial waves needed to achieve  convergence within 6 digits
    is about 10  for $E=0.25$ and is no more than 20 for $E=4.0$.
    At these relatively low energies,
    the use of local interpolation polynomials
    instead of the usual Legendre polynomials to treat the $x$-variable
    does not appear to give any
    computational advantage. In the context of
     a model nucleon-nucleon potential,
    Kessler, Payne and Polyzou [5] had reached a similar conclusion
     for the use of a wavelet basis  in the Galerkin method.
     Whether other bases
    (like global functions or more sophisticated
    local interpolation
    polynomials in conjunction with more elaborate
    discretization grids on the $q-x$ plane)  might lead to
    computational benefits at this energy range
    over Legendre basis remains to be
    explored. It is conceivable that to beat
    the Legendre-function representation of the $x$-variable,
    one  should use bases that are not of simple direct-product type,
    but  entangle  $q$ and $x$ variables via, perhaps, a suitable
    coordinate transformation.

\begin {center} {\bf REFERENCES  } \end{center}

\begin{enumerate}

    \item Ch. Elster, J. H. Thomas, and W. Glockle,
  Few-Body Syst. {\bf 24}, 55 (1998);\\
   D. Huber, W. Glockle, and
   A.  Bomelburg,
   Phys. Rev. C {\bf 42}, 2342 (1990).

\item W. Schadow, Ch. Elster,  and W. Glockle,
  Few-Body Syst.  {\bf 28}, 15 (2000).

\item J. Shertzer and A. Temkin, Phys. Rev. A {\bf 63}. 062714
(2001).

\item G. L. Caia, V. Pascalutsa and  L. E. Wright, Phys. Rev. C
{\bf 69}, 034003 (2004).

\item B. M. Kessler, G. L. Payne, and W. N. Polyzou, Phys. Rev. C
{\bf 70},  034003 (2004).

\item A. S. Kadyrov, I. Bray, A. T. Stelbovics, and B. Saha,
 J. Phys. B {\bf 38},  509 (2005).
\item G. Ramalho, A. Arriaga, and M. T. Pena,
 Few-Body Syst.  {\bf 39}, 123 (2006).

\item  M. Rodriguez-Gallardo, A. Deltuva,  E. Cravo,
  R. Crespo, and A. C. Fonseca,
Phys. Rev. C  {\bf 78},  034602 (2008).

\item  M. Rodriguez-Gallardo, A. Deltuva,
  R. Crespo, E. Cravo, and A. C. Fonseca,
Eur. Phys. J A  {\bf 42},  601 (2009).

\item A. S. Kadyrov, I. B. Abdurakhmanov, I. Bray,  and A. T.
Stelbovics, Phys. Rev. A  {\bf 80}, 022704 (2009).

\item   J. Golak,  R. Skibinski,  H. Witala,  K. Topolnicki,
 W. Glockle,  A. Nogga, and  H. Kamada,
  Few-Body Syst.  {\bf 53},  237(2012).

 \item S. Veerasamy, Ch. Elster,  and W. N. Polyzou,
 Few-Body Syst. (2012). doi:10.1007/s00601-012-0476-1

 \item H. Liu, Ch. Elster, and  W. Glockle,
 Phys. Rev. C  {\bf 72},  054003 (2005).

 \item Ch. Elster,  W. Glockle, and H. Witala,
   Few-Body Syst.  {\bf 45}, 1 (2009).

 \item K. E. Atkinson, {\it A Survey of Numerical Methods for the Solution of
 Fredholm Integral Equations of the Second Kind} (SIAM, Philedelphia, 1976)

 \item S. K. Adhikari, {\it Variational Principles and the Numerical Solution
  of Scattering Problems} ( Wiley, New York, 1998).

  \item B. A. Finlayson, {\it The Method of Weighted Residuals and
  Variational Principles} ( Academic Press, New York,  1972).

  \item C. A. J. Fletcher, {\it Computational Galerkin Methods }
  (Springer, New  York, 1984).

   \item P. M. Prenter, {\it Splines and Variational Methods}
    (Wiley, New York, 1975).

    \item J. P. Boyd, {\it Chebychev and Fourier Spectral Methods},
    2nd ed. (Dover, New York, 2001)

 \item C. D. Meyer, {\it Matrix Analysis and Applied Linear Algebra} (SIAM,
 Philedelphia, 2000).

 \item Y. Saad, {\it Iterative Methods for Sparse Linear Systems}, 2nd ed.
  (SIAM, Philedelphia, 2003).

  \item R. T. Behrens and L. L. Scharf, IEEE Trans. Signal Processing {\bf 42}, 1413 (1994).

  \item P. O. Lowdin, {\it Linear Algebra for Quantum Theory}
  (Wiley, New York, 1998).

 \item Z. C. Kuruoglu and D. A. Micha, J. Chem. Phys. {\bf 72}, 3328 (1980).

  \item    O. Christensen, {\it An Introduction to Frames and Riesz Bases}
   (Birkhauser, Boston, 2003).

 \item O. A. Rubtsova, V. N. Pomerantsev, and V. I. Kukulin,
    Phys. Rev. C  {\bf 79},  064602 (2009).

  \item Z. C. Kuruoglu and F. S. Levin, Phys. Rev. A {\bf 46}, 2304 (1992).

 \item Z. C. Kuruoglu and D. A. Micha, J. Chem. Phys. {\bf 80}, 4262 (1984).

\item J. Lund, J. and K. Bowers, K.  {\it Sinc methods for
quadrature and differential equations} (SIAM,
 Philedelphia, 1992).

\item    C. G. Broyden and M. T. Vespucci, {\it Krylov Solvers for
    Linear Algebraic Systems} (Elsevier, Amsterdam, 2004).

\end{enumerate}

\newpage

\noindent TABLE\ I. \\
\noindent { Dependence of the scattering amplitudes on the
momentum cutoff. Shown are the results of converged Nystrom
calculations for the scattering amplitude  $A(x;E)$ for
 $E=0.25$  and $E=4.0$.\\

\begin{tabular}{ccrrrr}
\hline \multicolumn{1}{c}{$\ \ E\ \  $}
  &\multicolumn{1}{c}{$\ \ q_{max} \ \ $}
  & \multicolumn{1}{c}{\ \ s-wave \ \ }
  &\multicolumn{1}{c}{$\ \ x=-1.0 \ \ $}
& \multicolumn{1}{c} { $\ \ x=0.0 \ \ $}
&\multicolumn{1}{c}{ \ \ $ x=1.0 \ \  $}\\
\hline \hline \multicolumn{2}{c}{\ } &
\multicolumn{4}{c}{Real part of scattering amplitude }\\
\cline{3-6}

 0.25  &   $30$ & 0.868839 & 0.725575  & 0.861919 & 1.040082 \\
\ & $> 1000$  &  0.868552 & 0.725289 & 0.861633 & 1.039795 \\
\\
4.0  &   $30$ & 0.245902 & 0.050810  & 0.162513
& 0.978586 \\
  & $> 1000$  & 0.245919 & 0.050828  & 0.162530
& 0.978604 \\
 \hline \hline
 \multicolumn{2}{c}{\ } &
\multicolumn{4}{c}{Imaginary part of scattering amplitude } \\
\cline{3-6}

   0.25  &   $30$  & 1.495095 & 1.495095  & 1.495086 & 1.499179 \\
     \      & $>1000$  & 1.495598 & 1.491553  & 1.495589 & 1.499682 \\
  \\
       4.0  &   $30$    & 0.204919 & 0.144448  & 0.197058 & 0.299073 \\
   \        & $> 1000$ & 0.205014 & 0.144543  & 0.197152 & 0.299168 \\
 \hline \hline
\end{tabular}

\newpage

\noindent TABLE\ II. Convergence with respect to the basis size in
the $q$-variable for the Schwinger variational method (SVM)
employing a direct-product basis of piecewise quadratic
polynomials. Shown are the scattering amplitudes $A(x;E)$ at
$E=0.25$.  Parameters $N$ and $M$ denote the number of basis
basis functions in $q$ and $x$ variables, respectively.\\

\begin{tabular}{ccllll}
\hline \hline
  \multicolumn{1}{c}{$\ \ N\ \ $ }
  & \multicolumn{1}{c}{$\ \ M\ \ $ }
  & \multicolumn{1}{l}{\ \ \  s-wave\ \ \ }
  & \multicolumn{1}{l}{\ \ \ x=-1.0 \ \  \  }
  & \multicolumn{1}{l} {\ \ \ x=0.0 \ \ \ }
&\multicolumn{1}{l}{\ \ \  x=1.0 \ \ \  } \\

\hline \hline
\multicolumn{2}{c}{ } & \multicolumn{4}{c} {Real part of scattering amplitude}\\
\cline{3-6}
  9 &  11 & 0.869427 & 0.726166  & 0.862508 & 1.040668\\
 17  & 11 & 0.868557 & 0.725306  & 0.861650 & 1.039812\\
 25  & 11 & 0.868554 & 0.725290  & 0.861635 & 1.039797 \\
 33  & 11 & 0.868552 & 0.725289  & 0.861633 & 1.039795 \\
 41  & 11 & 0.868552 & 0.725289  & 0.861633 & 1.039795\\
\hline
 \multicolumn{2}{l}{Nystrom}
  & \multicolumn{1}{l}{0.868552 }
  & \multicolumn{1}{l}{0.725289}
  & \multicolumn{1}{l}{0.861633}
  & \multicolumn{1}{l}{1.039795  }\\
\hline
 \multicolumn{2}{l}{Ref.[3]}
  & \multicolumn{1}{l}{0.869 }
  & \multicolumn{1}{l}{0.725}
  & \multicolumn{1}{l}{0.862}
  & \multicolumn{1}{l}{1.040  }\\
\hline \hline
\multicolumn{2}{c}{ } & \multicolumn{4}{c} {Imaginary part of scattering amplitude}\\
\cline{3-6}
 9   & 11 &  1.494061  &  1.490016 &  1.494051 &  1.498145 \\
 17  & 11 &  1.495568  &  1.491522 &  1.495558 &  1.499652 \\
 25  & 11 &  1.495595  &  1.491550 &  1.495586 &  1.499679 \\
 33  & 11 &  1.495598  &  1.491552 &  1.495588 &  1.499681 \\
 41  & 11 &  1.495598  &  1.491553 &  1.495588 &  1.499682 \\
\hline
 \multicolumn{2}{l}{Nystrom}
 & \multicolumn{1}{l}{1.495598 } &
\multicolumn{1}{l}{1.491553} & \multicolumn{1}{l}{1.495589}
& \multicolumn{1}{l}{1.499682}\\
\hline
 \multicolumn{2}{l}{Ref.[3]}
 & \multicolumn{1}{l}{1.495 } &
\multicolumn{1}{l}{1.491}
 & \multicolumn{1}{l}{1.496}
 & \multicolumn{1}{l}{1.500}\\
\hline \hline

\end{tabular}

\newpage

\noindent TABLE\ III. Convergence with respect to the basis size
 for the Schwinger variational method (SVM)
employing a direct-product basis of piecewise quadratic
polynomials. Shown are the scattering amplitudes $A(x;E)$ at
$E=4.0\, $.  \\

\begin{tabular}{ccllll}
\hline \hline
  \multicolumn{1}{c}{$\ \ N\ \ $ }
  & \multicolumn{1}{c}{$\ \ M\ \ $ }
  & \multicolumn{1}{l}{\ \ \  s-wave\ \ \ }
  & \multicolumn{1}{l}{\ \ \ x=-1.0 \ \  \  }
  & \multicolumn{1}{l} {\ \ \ x=0.0 \ \ \ }
&\multicolumn{1}{l}{\ \ \  x=1.0 \ \ \  } \\
\hline \hline
\multicolumn{2}{c}{ } & \multicolumn{4}{c} {Real part of scattering amplitude}\\
\cline{3-6}
 17  & 21 & 0.245910 & 0.0509076  & 0.162662 & 0.976805\\
 25  & 21 & 0.245919 & 0.0508297  & 0.162535 & 0.978505\\
 33  & 21 & 0.245919 & 0.0508287  & 0.162530 & 0.978584\\
     & 31 & 0.245919 & 0.0508275  & 0.162530 & 0.978600\\
 41  & 21 & 0.245919 & 0.0508287  & 0.162530 & 0.978588\\
     & 31 & 0.245919 & 0.0508276  & 0.162530 & 0.978604\\

\hline
 \multicolumn{2}{l}{Nystrom}
 & \multicolumn{1}{l}{0.245919}
  & \multicolumn{1}{l}{0.0508275}
& \multicolumn{1}{l}{0.162530}
& \multicolumn{1}{l}{0.978604  }\\
\hline
 \multicolumn{2}{l}{Ref.[3]}
 & \multicolumn{1}{l}{0.246}
 & \multicolumn{1}{l}{0.051}
 & \multicolumn{1}{l}{0.164}
& \multicolumn{1}{l}{0.979  }\\
\hline \hline
\multicolumn{2}{c}{ } & \multicolumn{4}{c} {Imaginary part of scattering amplitude}\\
\cline{3-6}
 17  & 21 & 0.204964 & 0.144588 & 0.197135 & 0.298868\\
 25  & 21 & 0.205013 & 0.144544  & 0.197152 & 0.299158\\
 33  & 21 & 0.205014 & 0.144543  & 0.197152 & 0.299168\\
     & 31 & 0.205014 & 0.144543  & 0.197152 & 0.299168\\
 41  & 21 & 0.205014 & 0.144543  & 0.197152 & 0.299168\\
     & 31 & 0.205014 & 0.144543  & 0.197152 & 0.299168\\
\hline
 \multicolumn{2}{l}{Nystrom}
  & \multicolumn{1}{l}{0.205014 }
  & \multicolumn{1}{l}{0.144543}
 & \multicolumn{1}{l}{0.197152}
 & \multicolumn{1}{l}{0.299168}\\
\hline
 \multicolumn{2}{l}{Ref. [3]}
 & \multicolumn{1}{l}{0.205 }
 & \multicolumn{1}{l}{0.145}
 & \multicolumn{1}{l}{0.197}
 & \multicolumn{1}{l}{0.300}\\
\hline \hline

\end{tabular}

\newpage

\noindent TABLE\  IV. \\
\noindent { Calculations with the Galerkin method with the
direct-product basis of piecewise quadratic polynomials. Shown are
the
results for scattering amplitude $A(x;E)$ at $E=0.25$. \\

\begin{tabular}{cclll}
\hline \hline
  \multicolumn{1}{c}{$\ \ N\ \ $ }
  & \multicolumn{1}{c}{$\ \ M\ \ $ }
  & \multicolumn{1}{l}{\ \ \ x=-1.0 \ \  \  }
  & \multicolumn{1}{l} {\ \ \ x=0.0 \ \ \ }
&\multicolumn{1}{l}{\ \ \  x=1.0 \ \ \  } \\
\hline \hline
\multicolumn{2}{c}{ } & \multicolumn{3}{l} {Real part of scattering amplitude}\\
\cline{3-5}
 9  & 11 & 0.729404   & 0.864266 & 1.042470\\
 17  & 11 & 0.724437   & 0.861607 & 1.040121\\
 25  & 11 & 0.725323   & 0.861596 & 1.039761\\
 33  & 11 & 0.725260   & 0.861655 & 1.039803 \\
 41  & 11 & 0.725295   & 0.861621 & 1.039770 \\
 49  & 11 & 0.725292   & 0.861640 & 1.039784\\
     & 15 & 0.725287   & 0.861640 & 1.039796 \\
     & 21 & 0.725285   & 0.861641 & 1.039801 \\
     & 25 & 0.725284   & 0.861641 & 1.039802 \\
 81  & 25 & 0.725289   & 0.861635 & 1.039795\\
\hline
 \multicolumn{2}{l}{Nystrom}
 & \multicolumn{1}{l}{0.725289}
 & \multicolumn{1}{l}{0.861633}
 & \multicolumn{1}{l}{1.039795  }\\
\hline \hline
\multicolumn{2}{c}{ } & \multicolumn{3}{l} {Imaginary part of scattering amplitude}\\
\cline{3-5}
 9  & 11 &  1.486512&  1.490502 &  1.494557 \\
 17  & 11 &  1.491686 &  1.495757 &  1.499885 \\
 25  & 11 &  1.491595 &  1.495629 &  1.499721 \\
 33  & 11 &  1.491527 &  1.495564 &  1.499660 \\
 41  & 11 &  1.491568 &  1.495604 &  1.499697 \\
 49  & 11 &  1.491542 &  1.495578 &  1.499672 \\
     & 15 &  1.491542 &  1.495578 &  1.499672 \\
     & 21 &  1.491542 &  1.495578 &  1.499672 \\
     & 25 &  1.491542 &  1.495578 &  1.499672\\
 81  & 25 &  1.491550 &  1.495586 &  1.499680 \\
\hline
 \multicolumn{2}{l}{Nystrom}
 &\multicolumn{1}{l}{1.491553}
 & \multicolumn{1}{l}{1.495589}
& \multicolumn{1}{l}{1.499682}\\
\hline \hline
\end{tabular}

\newpage

\noindent TABLE\  V. \\
\noindent { Calculations with the collocation method using the
direct-product basis of piecewise quadratic polynomials. Shown are
the results  for the scattering amplitude
 $A(x;E)$ at $E=0.25$. \\

\begin{tabular}{cclll}
\hline \hline
  \multicolumn{1}{c}{$\ \ N\ \ $ }
  & \multicolumn{1}{c}{$\ \ M\ \ $ }
  & \multicolumn{1}{l}{\ \ \ x=-1.0 \ \  \  }
  & \multicolumn{1}{l} {\ \ \ x=0.0 \ \ \ }
&\multicolumn{1}{l}{\ \ \  x=1.0 \ \ \  } \\
\hline \hline
\multicolumn{2}{c}{ } & \multicolumn{3}{c} {Real part of scattering amplitude}\\
\cline{3-5}
 9  & 11 & 0.723380   & 0.859676 & 1.037778\\
 17  & 11 & 0.725509   & 0.861862 & 1.040035\\
 25  & 11 & 0.725231   & 0.861573 & 1.039734\\
 33  & 11 & 0.725319   & 0.861665 & 1.039830 \\
 41  & 11 & 0.725276   & 0.861621 & 1.039784 \\
 49  & 11 & 0.725298   & 0.861643 & 1.039807\\
     & 21 & 0.725298   & 0.861642 & 1.039805 \\
     & 25 & 0.725298   & 0.861642 & 1.039805 \\
\hline
 \multicolumn{2}{l}{Nystrom}
 &  \multicolumn{1}{l}{0.725289}
 & \multicolumn{1}{l}{0.861633}
 & \multicolumn{1}{l}{1.039795  }\\
\hline \hline
\multicolumn{2}{c}{ } & \multicolumn{3}{c} {Imaginary part of scattering amplitude}\\
\cline{3-5}
 9  & 11 &  1.494976 &  1.499008 &  1.503099 \\
 17  & 11 &  1.491153 &  1.495189 &  1.499283 \\
 25  & 11 &  1.491659 &  1.495695 &  1.499788 \\
 33  & 11 &  1.491499 &  1.495535 &  1.499628 \\
 41  & 11 &  1.491576 &  1.495612 &  1.499705 \\
  49& 11 &  1.491537 &  1.495573 &  1.499667 \\
     & 21 &  1.491536 &  1.495572 &  1.499665 \\
     & 25 &  1.491536 &  1.495572 &  1.499665 \\
\hline
 \multicolumn{2}{l}{Nystrom}
 &\multicolumn{1}{l}{1.491553}
  & \multicolumn{1}{l}{1.495589}
 & \multicolumn{1}{l}{1.499682}\\
\hline \hline
\end{tabular}

\newpage

\noindent TABLE\  VI. \\
\noindent { Calculations with the hybrid-collocation method
 using piecewise quadratic polynomials for both $q$ and $x$ variables. Shown
are the results  for the scattering amplitude
 $A(x;E)$ at $E=0.25$. \\

\begin{tabular}{cclll}
\hline \hline
  \multicolumn{1}{c}{$\ \ N\ \ $ }
  & \multicolumn{1}{c}{$\ \ M\ \ $ }
  & \multicolumn{1}{l}{\ \ \ x=-1.0 \ \  \  }
  & \multicolumn{1}{l} {\ \ \ x=0.0 \ \ \ }
&\multicolumn{1}{l}{\ \ \  x=1.0 \ \ \  } \\
\hline \hline
\multicolumn{2}{c}{ } & \multicolumn{3}{c} {Real part of scattering amplitude}\\
\cline{3-5}
  9  & 11 & 0.726049  & 0.862408 & 1.040588\\
 17  & 11 & 0.725288   & 0.861632 & 1.039795\\
 25  & 11 & 0.725290   & 0.861534 & 1.039797\\
 33  & 11 & 0.725289   & 0.861633 & 1.039795 \\
 41  & 11 & 0.725289   & 0.861633 & 1.039795 \\

\hline
 \multicolumn{2}{l}{Nystrom}
& \multicolumn{1}{l}{0.725289} & \multicolumn{1}{l}{0.861633}
& \multicolumn{1}{l}{1.039795  }\\
\hline \hline
\multicolumn{2}{c}{ } & \multicolumn{3}{c} {Imaginary part of scattering amplitude}\\
\cline{3-5}
  9  & 11 &  1.490365 &  1.494374 &  1.498443 \\
 17  & 11 &  1.491553 &  1.495589 &  1.499683 \\
 25  & 11 &  1.491550 &  1.495586 &  1.499679 \\
 33  & 11 &  1.491552 &  1.495588 &  1.499682 \\
 41  & 11 &  1.491553 &  1.495589 &  1.499682 \\
\hline
 \multicolumn{2}{l}{Nystrom}
 &\multicolumn{1}{l}{1.491553}
 & \multicolumn{1}{l}{1.495589}
& \multicolumn{1}{l}{1.499682}\\
\hline \hline
\end{tabular}

\newpage

\noindent TABLE  VII.  Scattering amplitude
 $A(x;E)$ for $E=0.25$ calculated from
 Schwinger variational method using a basis constructed from
 piecewise quadratic  polynomials under the action of $G_0\ $.
 In these calculations $M=11$.\\

\begin{tabular}{cclll}
 \hline \hline
  \multicolumn{1}{c}{$\ \ N\ \ $ }
  & \multicolumn{1}{l}{\ \ \ x=-1.0 \ \  \  }
  & \multicolumn{1}{l} {\ \ \ x=0.0 \ \ \ }
&\multicolumn{1}{l}{\ \ \  x=1.0 \ \ \  } \\
\hline \hline
\multicolumn{1}{r}{ } & \multicolumn{3}{l} {Real part of scattering amplitude}\\
\cline{2-4}
    9      & 0.725287  & 0.861630   & 1.039792 \\
   17      & 0.725288   & 0.861632   & 1.039795 \\
   25      & 0.725288   & 0.861633   & 1.039795 \\
   \hline
  \multicolumn{1}{r}{ } & \multicolumn{3}{l} {Imaginary  part of scattering amplitude}\\
\cline{2-4}
 9    &   1.491554 &  1.495589 &  1.499681 \\
 17   &   1.491553 &  1.495589 &  1.499683 \\
 25   &   1.491553 &  1.495589 &  1.499682 \\
\hline \hline

\end{tabular}

\newpage

\noindent TABLE\  VIII. \\
\noindent { Calculations with SVM using $N$ Gaussians as the
$q$-basis. The $x$-basis consists of $M$  piecewise quadratic
polynomials. Shown are the results for the scattering amplitude
 $A(x;E)$ at $E=0.25$. \\

\begin{tabular}{cclll}
 \hline \hline
  \multicolumn{1}{c}{$\ \ N\ \ $ }
  & \multicolumn{1}{c}{$\ \ M\ \ $ }
  & \multicolumn{1}{l}{\ \ \ x=-1.0 \ \  \  }
  & \multicolumn{1}{l} {\ \ \ x=0.0 \ \ \ }
&\multicolumn{1}{l}{\ \ \  x=1.0 \ \ \  } \\
\hline \hline
\multicolumn{2}{c}{ } & \multicolumn{3}{c} {Real part of scattering amplitude}\\
\cline{3-5}
  9  & 11 & 0.725429  & 0.861771& 1.039932\\
 17  & 11 & 0.725298   & 0.861642 & 1.039805\\
 25  & 11 & 0.725292   & 0.861636 & 1.039798\\
 33  & 11 & 0.725280   & 0.861634 & 1.039797 \\
 41  & 11 & 0.725289   & 0.861633 & 1.039796 \\

\hline
 \multicolumn{2}{l}{Nystrom}
& \multicolumn{1}{l}{0.725289} & \multicolumn{1}{l}{0.861633}
& \multicolumn{1}{l}{1.039795  }\\
\hline \hline
\multicolumn{2}{c}{ } & \multicolumn{3}{c} {Imaginary part of scattering amplitude}\\
\cline{3-5}
  9  & 11 &  1.491310 &  1.495345 &  1.499439 \\
 17  & 11 &  1.491536 &  1.495572 &  1.499665 \\
 25  & 11 &  1.491547 &  1.495583 &  1.499676 \\
 33  & 11 &  1.491550 &  1.495586 &  1.499679 \\
 41  & 11 &  1.491552 &  1.495587 &  1.499681 \\
\hline
 \multicolumn{2}{l}{Nystrom}
 &\multicolumn{1}{l}{1.491553}
  & \multicolumn{1}{l}{1.495589}
 & \multicolumn{1}{l}{1.499682}\\
\hline \hline
\end{tabular}

\end {document}